\documentclass[conference]{IEEEtran}
\usepackage{cite}
\usepackage{amsmath}
\usepackage{amsthm}
\usepackage[moderate,title]{savetrees}
\usepackage{mathtools}
\usepackage{algorithmic}

\usepackage{array}

\usepackage{url}
\usepackage[inkscapeformat=png]{svg}
\usepackage{siunitx}
\usepackage{amssymb}
\usepackage[capitalise]{cleveref}
\usepackage{dsfont}
\usepackage{soul}
\usepackage{comment}
\usepackage{multirow}
\usepackage{lipsum}
\usepackage{derivative}
\usepackage{lettrine}
\usepackage{subcaption}
\usepackage{tikz}
\usepackage{float}
\usepackage{tabu}
\usepackage{placeins}
\usepackage[nodayofweek]{datetime}
\newdate{date}{14}{12}{2023}

\DeclareMathOperator{\sgn}{sgn}
\newtheorem{prob}{Problem}

\begin{document}

\title{\Huge{\textnormal{Multi-layer optimisation of hybrid energy storage systems for electric vehicles}}}

\author{Wouter Andriesse, Jorn van Kampen, and Theo Hofman}% <-this % stops a space

% make the title area
\maketitle
\IEEEpeerreviewmaketitle
%% As per your sections
\begin{abstract}
    This research presents a multi-layer optimization framework for hybrid energy storage systems (HESS) for passenger electric vehicles to increase the battery system's performance by combining multiple cell chemistries. Specifically, we devise a battery model capturing voltage dynamics, temperature and lifetime degradation solely using data from manufacturer datasheets, and jointly optimize the capacity distribution between the two batteries and the power split, for a given drive cycle and HESS topology. The results show that the lowest energy consumption is obtained with a hybrid solution consisting of a NCA-NMC combination, since this provides the best trade-off between efficiency and added weight. 

    % he results show that although a single-cell battery results in the lowest TCO, the lowest energy consumption is obtained with a hybrid solution. 
    
    % The goal of the research is to explore the possibility of increasing battery performance by combining two battery cell chemistries. 
    % We introduce a modeling framework that evaluates the battery's total cost of ownership (TCO) and energy consumption over a predefined drive cycle for a given HESS topology. 
    % We introduce a HESS topology and three different cell chemistry combinations.
    
    % We define a two-layer optimization problem consisting of the optimal battery sizing problem, where we optimize the capacity distribution between the high-energy and high-power battery, and the optimal control problem, where we optimize the power split over the drive cycle. 
    % The initial results show that a single-cell battery constructed of a cell with a low production cost and low capacity degradation has still the lowest TCO. The results for optimization to solely minimal energy consumption show that a hybrid combination of high-energy and high-power reduces potentially energy consumption by trading battery weight and efficiency.
\end{abstract}
\section{Introduction}
% %\lettrine[lines=3]{D}{ue} to the ongoing climate crisis \cite{Lindsey2020}, all of humanity is urged to reduce their environmental impact. 
% The transport sector is traditionally powered by carbon fossil fuels \cite{U.S.EnergyInformationAdministrationEIA2021}, which have a tremendous environmental impact. 
% To mitigate this pollution, alternative power sources must be explored to reduce this pollution. Electrification and electrochemical energy storage in battery packs are popular alternatives since the battery packs can be charged from energy sources with a low environmental impact. This increase in energy demand from the electrical network \cite{EuropeanEnvironmentAgency2016} will result in new problems concerning electrical grid management. The combination of these grid challenges with the high cost and significant environmental impact of lithium batteries \cite{OLIVEIRA2015354} leads to the desire for ever higher efficiency levels of battery electric vehicles. 
% \par
During the last few decades, significant effort has been put into improving lithium-ion battery cells \cite{YOSHINO20141} by reducing cost and enhancing battery cell characteristics, like energy and power density. This resulted in both faster charging times and longer ranges of battery electric vehicles (BEV). However, current lithium-ion cells always show a trade-off between energy and power density, resulting in limited range when faster charge times are desired or vice versa. To mitigate this trade-off, a hybrid energy storage systems (HESS) consisting of two battery packs with different chemistries could be employed. Hereby, the HESS is a combination of a high-energy and high-power battery that are linked through a voltage balancing device, thereby combining the strengths of both the individual cells. To investigate the potential benefits, algorithms are required to optimize both the relative sizing and the energy flow between the two batteries.
%
% Problem is that single cells always make a trade-off in power and energy
% Can we do better by combining their strengths?
% To investigate this we need algorithms to optimally size and control the two batteries. 
%
% Another way to improve battery performance is to reduce overall vehicle consumption and achieve the same performance with less energy content. Much research is focused on reducing the losses in the motor, power electronics and drivetrain components.
%
% Our research will focus on improving battery efficiency by introducing a Hybrid Energy Storage System (HESS) consisting of two battery packs. One of the battery packs is a high-energy type, constructed of battery cells with high energy density, and the other battery is a high-power type, constructed of battery cells with high power density. Our research aims to investigate the potential benefits of using a hybrid energy storage system (HESS) compared to single-cell battery design for private electric vehicles.
\subsubsection*{Related literature}
Previous studies regarding HESSs focused on the topology, as well as the types of energy sources to combine. In \cite{Zimmermann2016}, the authors present a summary of a large set of HESS implementations and evaluate them for electric vehicles. They introduce three classes of HESS topologies: passive, active and discrete HESS. 
They consider the HESS on a topology level, but disregard battery sizing due to the qualitative approach of the research.
\par
Furthermore, much research around HESSs for electric vehicles focuses on combining batteries with supercapacitors \cite{hesssupercap.,hesscap1,hesscap2}. The research in \cite{hesscap2} also considers other HESS topologies, where a battery pack is combined with a flywheel, compressed air and magnetic energy storage. However, all these solutions have an inherent problem with storing energy for extended periods due to self-discharge or losses. This limits the usability of these energy storage systems to low-capacity and short-term storage, making them less usable for BEV. Conversely, rechargeable lithium-ion batteries have the advantage of being able to store a large amount of energy without losing that energy. 
\par 
In \cite{RaceHess}, the authors consider a HESS with rechargeable lithium battery packs for electric race cars. They explore a multi-layer optimization of the HESS with time minimization as the objective instead of energy or TCO. In addition, to obtain a tractable problem, the battery model is reduced in complexity. 
\par
In conclusion, to the best of the authors' knowledge, there are no studies jointly optimizing the relative sizing and control of a HESS, whilst accounting for secondary effects, such as battery lifetime and thermal effects.
 % The effect of the time minimization objective is that they require a complex vehicle model. As a result, the authors decide to reduce the complexity of the battery model to avoid a computationally heavy optimization problem. We consider a driving scenario with lower lateral accelerations and thus are able to reduce the complexity of the vehicle while having a more considerate battery model.         
\begin{figure}[t]
\centering
\includegraphics[width=0.85\columnwidth]{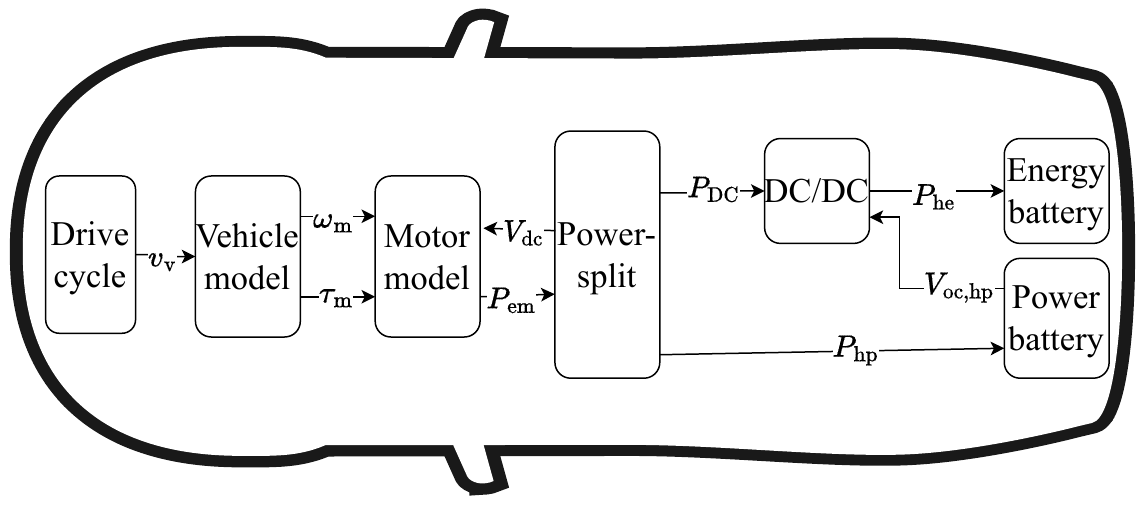}
\caption{Schematic overview of the vehicle model and HESS topology. The DC-DC converter is connected to the energy battery, and acts as the main connecting element between the two batteries and controls the power-split. The power battery is connected in parallel to the DC-DC, thereby determining the bus voltage.}
\label{fig:Vehicle Model}
\end{figure}
\subsubsection*{Statement of contributions}
This work presents an optimization framework to jointly optimize the relative sizing and power-split control between the two batteries in the HESS. The HESS topology consists of a high-energy and high-power battery that are connected through a single DC-DC converter, as shown in Fig.~\ref{fig:Vehicle Model}. Specifically, we devise a dynamic battery model that includes thermal behavior and capacity degradation as a function of the scaling parameters and the battery cell chemistry. Thereby, the cell chemistry influences various characteristics like cost, degradation and dynamic voltage response. In total, we consider four distinct chemistries, being Lithium Nickel-Cobalt-Aluminum Oxide (NCA), Lithium Nickel-Manganese-Cobalt Oxide (NMC), Lithium Iron Phosphate (LFP) and Lithium Titanium Oxide (LTO), which are commonly used in the automotive industry\cite{cellchems}. Finally, we explore the potential of a HESS in  the case where we want to minimize the energy consumption.

% Devise battery model
% different chemistries with characteristics
% Explore different cases of TCO and min energy.

% We explore the benefits of an active HESS constructed with lithium-ion battery packs and a single DC-DC converter connected to the high-energy battery. Based on two optimization goals: one where we want to minimize the total cost of ownership and one where we want to minimize energy consumption for a passenger electric vehicle. In the total cost of ownership, we consider the production costs, energy costs and battery replacement costs due to capacity degradation. We explore the optimal relative sizing and control for battery chemistry combinations. In this research, we consider a set of four different battery cell chemistries. Three of those are the most commonly used battery chemistries in automotive \cite{cellchems}: NCA, NMC and LFP batteries. The fourth cell chemistry we consider is an LTO battery cell \cite{ltochem} due to the low internal resistance, high discharge rate and low capacity degradation rate of this battery chemistry. 
\subsubsection*{Organization}
The outline of our paper is structured as follows: \cref{ch:methodology} defines the models we use to validate the HESS and single-cell batteries, together with the optimization framework. In \cref{ch:discussion}, we discuss the boundaries of our research, and in \cref{ch:resulst}, we discuss and analyze the results generated with the use case. \cref{ch:conclusion} states the outcomes of our    
research and offers an outlook on future research. 
% [\textit{Note: dear reviewer, the current draft paper has an excess length of about seven pages and will be significantly reduced for the final version.}]
\section{Methodology}\label{ch:methodology}
\begin{figure}[t]
\centering
\includegraphics[width=0.85\columnwidth]{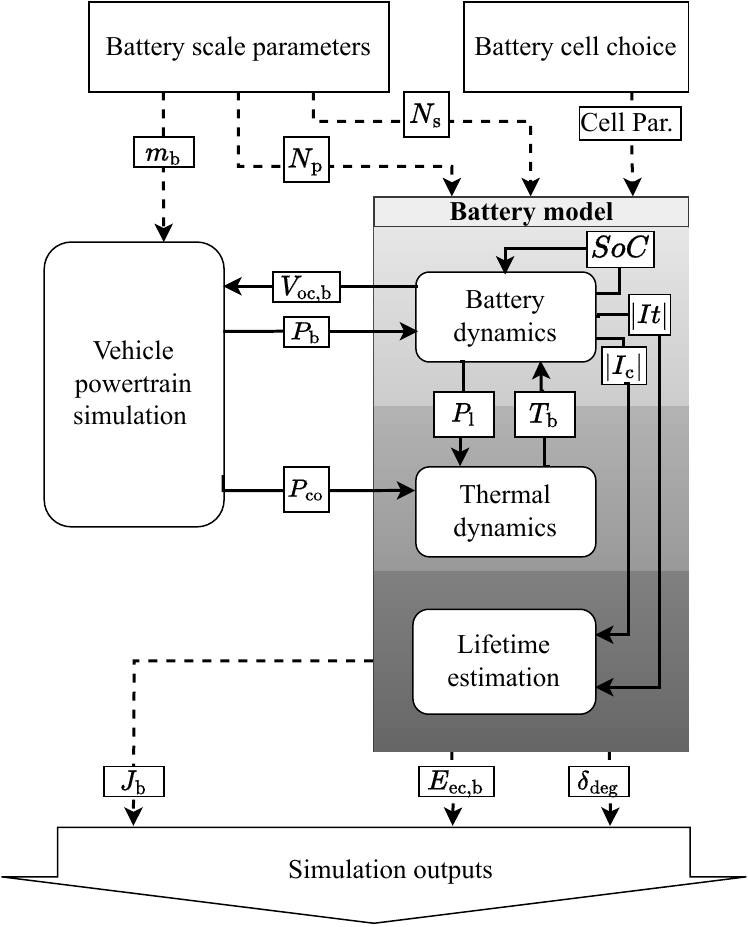}
\caption{Overview of the battery model structure. Given the battery scaling parameters and chemistry dependent cell characteristics, the model outputs energy consumption, cell degradation and cost. We include several dynamics, such as the voltage response, temperature and degradation.}
\label{fig:3LayerStructure}
\end{figure}

This section presents the battery model structure and optimization problem formulation. The battery model includes dynamic behavior, thermal dynamics and lifetime estimation in order to estimate production, operation and recycling costs. To optimize the relative sizing, we set the number of cells in series and parallel, and the battery mass as scalable parameters. An overview of the complete battery model is presented in Fig.~\ref{fig:3LayerStructure}. For the vehicle model, we use a quasi-static backward approach \cite{Guzellabook} for a given drive cycle. 
% the methodology, outlining the models describing the battery behavior, the vehicle, and the optimization problem containing the design and control problem. 
\subsection{Model structure}
We consider an active HESS topology with a bidirectional DC-DC converter connected to the high-energy battery. The high-power battery is connected in parallel with the DC-DC to the inverter. We consider this topology since it enables complete control over the power-split without the added complexity of two DC-DC converters.
By connecting the DC-DC converter to the energy battery, we can keep it compact, since its size scales with the power of the connected battery. 
\par
% From the predefined velocity profile and the total battery mass $m_\mathrm{b}$, we can calculate the battery power $P_\mathrm{b}$ which serves as an input to the battery model.

% The drive cycle predefines the vehicle speed $v_\mathrm{v}$. The vehicle model incorporates a road-load and gearbox model and computes the motor's shaft speed $\omega_\mathrm{em}$ and shaft torque $\tau_\mathrm{em}$. The motor model determines the electrical motor power $P_\mathrm{em}$ based on a motor-inverter efficiency map that depends on the shaft torque $\tau_\mathrm{em}$, shaft speed $\omega_\mathrm{em}$ and inverter bus/battery bus voltage $V_\mathrm{dc}$.
The power-split defines the bus voltage $V_\mathrm{dc}$, which is, in turn, influenced by the open-circuit voltage $V_\mathrm{oc,hp}$. We model the voltage dynamics as a function of the battery state-of-charge $SoC$, battery power $P_\mathrm{b}$ and battery temperature $T_\mathrm{b}$. As outputs, we obtain the battery cost $J_\mathrm{b}$, battery energy $E_\mathrm{ec,b}$ and normalized battery degradation $\delta_\mathrm{deg}$ to evaluate the battery performance. 

% The power-split and DC-DC converter define the bus voltage $V_\mathrm{dc}$ subjective to the power battery's open-circuit voltage $V_\mathrm{oc,hp}$. The power-split depends on the DC-DC bus voltage $V_\mathrm{dc}$, and it determines the power distribution between the high-energy battery and the high-power battery. The power-split is defined by solving the optimal control problem.
% \par 
% We model the high-power and high-energy battery independently according to the framework introduced in \cref{fig:3LayerStructure}. 
% The battery scale parameters and battery cell choice define the battery pack design. The cell parameters (Cell par.) define the pack's behavior on a cell level.
% $N_\mathrm{p}$ and $N_\mathrm{s}$ are integers that describe the number of cells connected in parallel ($N_\mathrm{p}$) and in series ($N_\mathrm{s}$). The battery mass $m_\mathrm{b}$ is used in the vehicle
% model to define the road load. The battery cost $J_\mathrm{b}$, battery energy $E_\mathrm{ec,b}$ and normalized battery degradation $\delta_\mathrm{deg}$ are parameters used to evaluate the battery performance. 
% The battery model consists of the battery dynamics, the thermal dynamics, and the lifetime estimation model. The battery dynamics model defines open-circuit voltage $V_\mathrm{oc,b}$ and battery losses $P_\mathrm{l}$ according to the power request $P_\mathrm{b}$, battery temperature $T_\mathrm{b}$ and the State of Charge $SoC$. The thermal dynamics model defines the battery temperature $T_\mathrm{b}$ as a function of the battery losses $P_\mathrm{l}$ and the heat dissipation $P_\mathrm{co}$. The lifetime estimation model determines the normalized battery degradation $\delta_\mathrm{deg}$ as a function of the absolute cell current $|I_\mathrm{c}|$ and the total capacity throughput $|It|$.  

\subsection{Battery dynamics}\label{ch:BatteryDynamics}
\begin{figure}[t]
\centering
\includegraphics[width=\columnwidth]{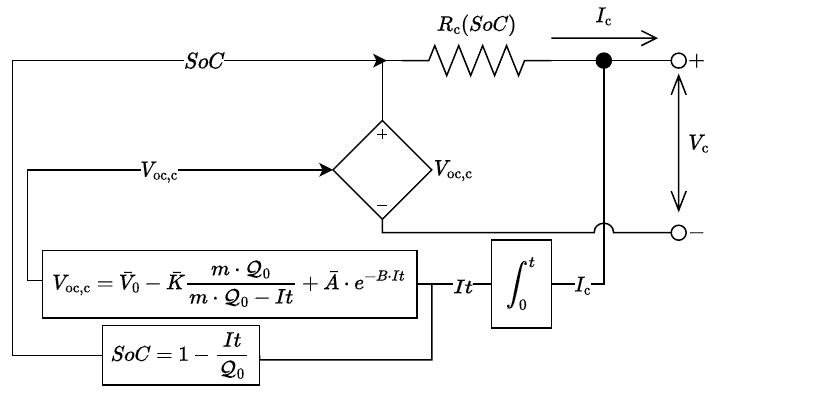}
\caption{Battery cell dynamics model without temperature dependency. Both the internal resistance and the open-circuit voltage are dependent on the state of charge.}
\label{fig:R-int}
\end{figure}

The battery dynamics model simulates the voltage response on a cell level \(V_{\mathrm{c}}\) and losses \(P_{\mathrm{l}}\) dependent on the cell temperature \(T_{\mathrm{c}}\), the load current \(I_{\mathrm{c}}\), and total discharged capacity \(It\), defined as the integral of current over time.
We model the battery behavior on a cell level and linearly scale it to a pack level.
For the cell parameter identification, we use commercially available datasheets. Given the limited information on such sheets, we choose a zeroth order equivalent circuit as shown in Fig~\ref{fig:R-int}, known as a Rint model~\cite{Tremblay2007}, extended to enable parameter extraction~\cite{D.SongGenericbatterymodel}. 
% \cref{fig:R-int} visually describes battery dynamics on the model on a cell level
% \begin{comment}
% The Rint model describes the cell voltage according to \cref{eq:Basic Rint V} and the cell losses according to \cref{eq:Basic Rint Loss}.
% \begin{align}
% V_{\mathrm{c}}=V_{\mathrm{oc,c}}-{R_{\mathrm{c}}} \cdot I_{\mathrm{c}}\label{eq:Basic Rint V}\\
% P_{\mathrm{l,c}}=I_{\mathrm{c}}^2\cdot {R_{\mathrm{c}}}
% \label{eq:Basic Rint Loss}
% \end{align}
% The open-circuit voltage $ V_{\mathrm{oc,c}}$ describes the cell voltage as a function of the State of Charge $SoC$ when no load is applied. 
% \cref{eq:Basic Rint V_oc} describes the open-circuit voltage according to the discharged capacity $It$, which is equivalent to the $SoC=1-\frac{It}{\mathcal{Q}_0}$, with $\mathcal{Q}_0$ the battery capacity.
% \begin{align}
%     V_{\mathrm{oc,c}}=\bar{V}_0+\bar{K}\cdot \frac{m\cdot \mathcal{Q}_0}{m\cdot \mathcal{Q}_0-It}+\bar{A}\cdot e^{-B \cdot It}
%     \label{eq:Basic Rint V_oc}
% \end{align}
% $\bar{V}_0$, $\bar{K}$, $m$, $\bar{A}$, $B$ are constants we extract from data according to the method described in \cite{D.SongGenericbatterymodel}.
% We consider the resistance to be dependent on the $SoC$. The cell resistance $R_\mathrm{c}(SoC)$ is modeled as a lookup table from data.
% \end{comment}

% \subsubsection{Temperature adjustment}
To model the influence of temperature on the voltage $V_{\mathrm{oc,c}}$ and resistance $R_\mathrm{c}$, we apply the extension of~\cite{D.SongGenericbatterymodel} to obtain
\begin{align}
\begin{split}
    V_{\mathrm{oc,c}}(SoC,T)=V_0(T)+K(T)\cdot \frac{m\cdot \mathcal{Q}(T)}{m\cdot \mathcal{Q}(T)-It}&\\+A(T)\cdot e^{-B \cdot It},&\label{eq:V_oc T}
\end{split}\\[1ex]
    V_\mathrm{c}(SoC,I,T)=V_{\mathrm{oc,c}}(SoC,T)-R_{\mathrm{c}}(SoC,T)\cdot I_\mathrm{c},& \label{eq:V_c T}
\end{align}
where $K$ is the polarization constant, $Q$ is the cell capacity, and $A$, $B$ and $V_\mathrm{0}$ are voltage constant.

\subsection{Thermal dynamics}\label{ch:Thermal model}
To determine the battery temperature, a thermal dynamics model is introduced in which we consider the battery cell and, by extension, the battery as a lumped mass. The thermal dynamics model of the cell is subject to the heat balance equation:
\begin{align}\label{eq:heatbalance}
     m_\mathrm{c}\cdot c_\mathrm{p,c}\cdot  \frac{\mathrm{d}T_\mathrm{c}}{\mathrm{d}t}=P_{\mathrm{l,c}}-P_\mathrm{co},
\end{align}
where $P_\mathrm{co}$ is the heat dissipation, $m_\mathrm{c}$ is the cell mass, and $c_\mathrm{p,c}$ is the specific heat capacity.
We consider liquid cooling at ambient temperature $T_\mathrm{amb}$ and assume exclusive heat transfer between the fluid and the cells.

We define the heat dissipated $P_\mathrm{co}$ from the battery cell according to 
\begin{align}
    P_\mathrm{co}={\kappa}_\mathrm{tot}\cdot (T_\mathrm{c}-T_\mathrm{amb}),\label{eq:Cooling power} 
\end{align}
  where $\kappa_\mathrm{tot}$ is the total heat transfer coefficient.
The battery pack temperature evolves uniformly since we do not consider thermal interaction between the battery cells. When scaling to a HESS, we compute the temperature separately for both battery packs due to the difference in cell parameters. 

\subsection{Ageing Model}\label{ch:ageing model}
To evaluate the degradation of the battery, we introduce an aging model. We model the cyclic degradation influenced by the magnitude of the current $|I_\mathrm{c}|$ and the total absolute capacity throughput $|It|$~\cite{Najera2023}
\begin{align}\label{eq:capdegtimestat}
    Q_\mathrm{deg}=a_\mathrm{cy}\cdot e^{|I_\mathrm{c}|\cdot b_\mathrm{cy}}\cdot |It|.
\end{align}
where $Q_\mathrm{deg}$ is the battery capacity degradation. $a_\mathrm{cy}$ and $b_\mathrm{cy}$ are constants subject to identification. 
Reformulating this as a function of distance travelled and normalizing by the EoL capacity, we obtain\begin{align} \label{eq:capdegdisdyn2}
\dot{\delta}_\mathrm{deg}=\dfrac{a_\mathrm{cy}\cdot e^{|I|\cdot b_\mathrm{cy}}\cdot |I|}{(1-SoH_\mathrm{EoL})\cdot \mathcal{Q}_0},
\end{align}
where $\dot{\delta}_\mathrm{deg}$ is the normalized degradation over distance.

\subsection{Power-split}\label{ch:hybrid model}
We can influence the power distribution between the high-power battery, denoted with subscript  $_\mathrm{hp}$, and the high-energy battery, denoted with subscript $_\mathrm{he}$, by setting the DC-DC converter voltage $V_\mathrm{dc}$ as
\begin{align}
    P_\mathrm{hp}=&\frac{V_\mathrm{oc,hp}-V_\mathrm{dc}}{R_\mathrm{hp}}\cdot V_\mathrm{dc}\label{eq:powerbatsplit}.
    % P_\mathrm{dc}=&P_\mathrm{em}-P_\mathrm{hp}\label{eq:dcbatsplit},\\
    % P_\mathrm{he}=&P_\mathrm{dc}\cdot \eta_\mathrm{dc}^{\sgn(P_\mathrm{dc})},\label{eq:energybatsplit}
    \end{align}  
We redefine the control parameter as the DC-DC converter power $P_\mathrm{dc}$ to obtain
\begin{align}
    P_\mathrm{hp}=&P_\mathrm{em}-P_\mathrm{dc}\label{eq:powerbatsplitdc},\\
    P_\mathrm{he}=&P_\mathrm{dc}\cdot \eta_\mathrm{dc}^{\sgn(P_\mathrm{dc})}\label{eq:energybatsplitdc},
\end{align}
where $\eta_\mathrm{dc}$ is the efficiency of the DC-DC converter.  
Then using~\eqref{eq:powerbatsplit}, the DC-DC voltage and power are related as
\medmuskip=0.6\medmuskip
    \begin{align}\label{eq:VDC}
    V_\mathrm{dc}&=\frac{1}{2}\cdot\left(V_\mathrm{oc,hp}+\sqrt{V_\mathrm{oc,hp}^2-4\cdot (P_\mathrm{em}-P_\mathrm{dc})\cdot R_\mathrm{hp}}\right).
\end{align}
\medmuskip=1.66666\medmuskip
% From the battery powers $P_\mathrm{hp}$ and $P_\mathrm{he}$, we can determine the energy extracted from the batteries at the terminals according to:
% \begin{align}
%     E_\mathrm{t,b}=\int_{0}^{t_\mathrm{f}} P_\mathrm{b}\;\mathrm{d}t.
% \end{align}

\subsection{Optimization problem}\label{ch:OptimisationProblem}
\begin{figure}[t]
\centering
\includegraphics[width=0.95\columnwidth]{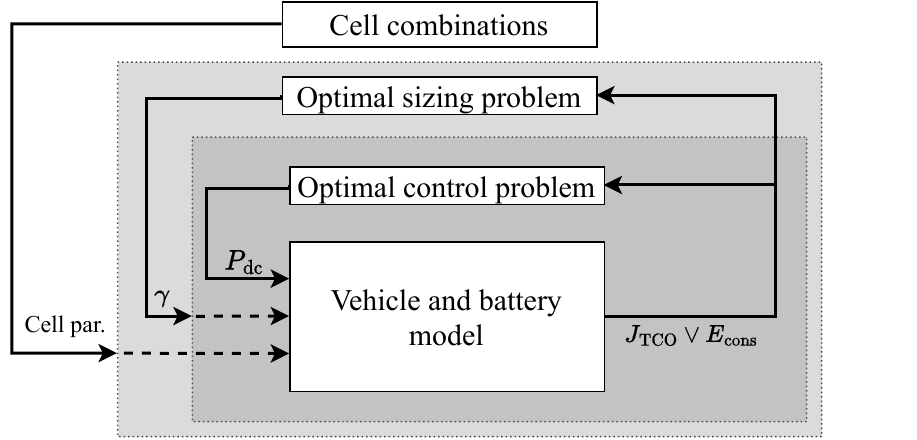}
\caption{Optimization layer scheme, where the inner layer solves the power-split between the two batteries and the outer layers determine the relative battery sizing and cell chemistries.}
\label{fig:optionion}
\end{figure}

In this research, we identify two possible optimization goals: one, where we pursue minimal energy consumption $J_\mathrm{E}$ and another, where we pursue minimal total cost of ownership $J_\mathrm{TCO}$ over a specified vehicle lifetime distance.
The optimization problem has two decision layers: battery sizing based on sizing parameter $\gamma$, and power-split control based on control parameter $P_\mathrm{dc}$. Following the optimization layers, we also explore the battery cell choice, but considering the limited choice and quick evaluation time, we use a brute-force approach. We consider three cell chemistry combinations: the NCA-NMC combination, the NCA-LFP combination and the NCA-LTO combination. In \cref{fig:optionion}, we present a graphic illustration of the optimization structure, where the objective functions are defined as follows:
\begin{align}
\begin{split}
J_\mathrm{TCO}=&\quad\:\mathcal{J}_{\mathrm{b,hp}}\cdot\left(d_\mathrm{l}\cdot\frac{\delta_\mathrm{deg,hp}}{d_\mathrm{c}}+1\right)\\&+\mathcal{J}_{\mathrm{b,he}}\cdot\left(d_\mathrm{l}\cdot\frac{\delta_\mathrm{deg,he}}{d_\mathrm{c}}+1\right)\\&+\mathcal{J}_{\mathrm{Q}}\cdot\frac{E_{\mathrm{ec,hp}}+E_{\mathrm{ec,he}}}{d_{\mathrm{c}}}\cdot d_\mathrm{l}, \label{eq:TCO}
\end{split}\\[1ex]
J_\mathrm{E}=&\quad\:E_{\mathrm{ec,hp}}+E_{\mathrm{ec,he}},\label{eq:Energy consumption}
\end{align}
where $\mathcal{J}_{\mathrm{b,hp}}$ and $\mathcal{J}_{\mathrm{b,he}}$ are battery costs, $\mathcal{J}_{\mathrm{Q}}$ is the energy cost, and $d_\mathrm{l}$ is the total lifetime in kilometers. 

% \begin{table}[t]
% \caption{Cost Function parameters}\label{tb:cost param}
% \centering
% {\tabulinesep=0.5mm
% \begin{tabu}{l|l|l|l}
% \hline
% Symbol & variable& value& Unit\\ \hline
% $\mathcal{J}_{\mathrm{Q}}$&energy cost \cite{energycost}& $0.32$& $\mathrm{\frac{\$}{kWh}}$\\
% $d_\mathrm{l}$&vehicle lifetime distance& 100000&$\mathrm{km}$ 
% \end{tabu}}
% \end{table}
The total electrochemical energies used by the batteries, $E_{\mathrm{ec,hp}}$ and $E_{\mathrm{ec,he}}$, are defined by 
\begin{align}\label{eq:Econs}
E_\mathrm{ec,b}=\int^{t_\mathrm{f}}_{0} V_\mathrm{oc,b}\cdot I_\mathrm{b} \;\mathrm{d}t.
\end{align}

\subsubsection{Optimal power-split}
The innermost optimization layer is the power-split control. The goal of the optimization is to find the optimal set of control inputs $P_\mathrm{dc}$ given the relative scaling of the two batteries $\gamma$, where the objective function, $J_\mathrm{TCO}$ or $J_\mathrm{E}$, is minimal within a set of constraints. The control problem with constraints is presented in \cref{prob:control P}. 
\begin{prob}[Power-split control]\label{prob:control P}
The power-split between the two batteries is the solution of
\begin{subequations}\label{eq:optimisationconstraints}  
\begin{align}
\mathbf{PSC}(\gamma)\  \coloneqq \  & \arg\min_{P_\mathrm{dc}} & & J_\mathrm{TCO}(P_\mathrm{dc},\gamma)\:\vee\: J_\mathrm{E}(P_\mathrm{dc},\gamma)  &\nonumber\\
& \textnormal{s.t. }  & & V_\mathrm{hp,min}\leq V_\mathrm{hp}\leq V_\mathrm{hp,max},&\\
& & & V_\mathrm{he,min}\leq V_\mathrm{he}\leq V_\mathrm{he,max},&\\
& & & I_\mathrm{hp,min}\leq I_\mathrm{hp}\leq I_\mathrm{hp,max},&\\
& & & I_\mathrm{he,min}\leq I_\mathrm{he}\leq I_\mathrm{he,max},&\\
& & &\labelcref{eq:V_oc T}-\labelcref{eq:VDC}.\nonumber&
\end{align}
\end{subequations}
\end{prob}
 We extract the battery level constraints for voltages and currents $V_\mathrm{b,max}$, $V_\mathrm{b,min}$, $I_\mathrm{b,max}$ and $I_\mathrm{b,min}$ from  datasheets without considering temperature effects.
 
 We rewrite the local constraints of \cref{prob:control P} to the control parameter $u(t)=P_\mathrm{dc}$ according to
\begin{subequations}\label{eq:PDCconstraints}
\begin{align}
    P_{\mathrm{dc}}\leq&  I_{\mathrm{he,max}}\cdot V_\mathrm{he} \cdot \eta_{\mathrm{dc}}^{-1},\\
    P_{\mathrm{dc}}\leq& \frac{V_{\mathrm{oc,he}}-V_{\mathrm{he,min}}}{{R_\mathrm{he}}\cdot \eta_{\mathrm{dc}}}\cdot V_{\mathrm{he,min}},\\ 
    P_{\mathrm{dc}}\leq& P_{\mathrm{em}}-(I_\mathrm{hp,{min}}\cdot V_\mathrm{hp}),\\
    P_{\mathrm{dc}}\leq& P_{\mathrm{em}}-\frac{V_{\mathrm{oc,hp}}-V_{\mathrm{hp,max}}}{{R_\mathrm{he}}}\cdot V_\mathrm{hp,{max}},\\
    P_{\mathrm{dc}}\geq&  I_\mathrm{he,{min}}\cdot V_\mathrm{he} \cdot \eta_{\mathrm{dc}},\\
    P_{\mathrm{dc}}\geq& \frac{V_{\mathrm{oc,he}}-V_\mathrm{he,{max}}}{{R_\mathrm{he}}}\cdot V_\mathrm{he,{max}}\cdot \eta_{\mathrm{dc}},\\
    P_{\mathrm{dc}}\geq&  P_{\mathrm{em}}-(I_\mathrm{hp,{max}}\cdot V_\mathrm{hp}),\\
    P_{\mathrm{dc}}\geq&  P_{\mathrm{em}}-\frac{V_{\mathrm{oc,hp}}-V_\mathrm{hp,{min}}}{{R_\mathrm{hp}}}\cdot V_\mathrm{hp,{min}},
\end{align}
\end{subequations}
to obtain the feasible domain $\mathcal{U}$ for control parameter $u(t)=P_\mathrm{dc}$.
We use Pontryagin's Minimum Principle (PMP) to solve the control problem with consideration of the system states while remaining computationally efficient compared to other optimal control strategies, such as dynamic programming.
The system dynamics are defined as
    \begin{align}\label{eq:xdot1}
        \dot{x}(t)=\begin{bmatrix}
             I_\mathrm{he}(t)/\mathcal{Q}_\mathrm{he} \\
             I_\mathrm{hp}(t)/\mathcal{Q}_\mathrm{hp} 
             \end{bmatrix}.
    \end{align}
   To define the system dynamics as a function of the system states and the control parameter, we take the power of each battery after the DC-DC converter
    \begin{align} 
    \begin{split}\label{eq:P_Dctocurrent}
        P_\mathrm{dc}(t)&=\eta_\mathrm{dc}\cdot(V_\mathrm{oc,he}(x_\mathrm{1})\cdot I_\mathrm{he}(t)-R_\mathrm{he}(x_1)\cdot I_\mathrm{he}^2(t))\\
        &={u}(t),
    \end{split}\\[1ex]
\begin{split}\label{eq:P_powertocurrent}
         P_\mathrm{hp}(t)&=V_\mathrm{oc,hp}(x_\mathrm{2})\cdot I_\mathrm{hp}(t)-R_\mathrm{hp}(x_2)\cdot I_\mathrm{hp}^2(t)\\&=P_\mathrm{em}(t)-P_\mathrm{dc}(t)=P_\mathrm{em}(t)-u(t),
\end{split}
\end{align}
and extract the current to include them in the dynamics
\begin{align} 
    \label{eq:xdot2}
    \renewcommand{\arraystretch}{2.5}
    \resizebox{0.89\columnwidth}{!}{$
        \dot{x}(t)\!=\!\begin{bmatrix}
            \frac{V_\mathrm{oc,he}(x_\mathrm{1})-\sqrt{V_\mathrm{oc,he}(x_\mathrm{1})^2-4\cdot u(t)\cdot\eta_\mathrm{dc}^{-1}\cdot R_\mathrm{he}(x_\mathrm{1})}}{\mathcal{Q}_\mathrm{he}\cdot2\cdot R_\mathrm{he}(x_1)}\\
            \frac{V_\mathrm{oc,hp}(x_\mathrm{2})-\sqrt{V_\mathrm{oc,hp}(x_\mathrm{2})^2-4\cdot(P_\mathrm{em}(t)-u(t))\cdot R_\mathrm{hp}\!(x_\mathrm{2})}}{\mathcal{Q}_\mathrm{hp}\cdot2\cdot R_\mathrm{hp}(x_2)}
             \end{bmatrix}$}.
\end{align}
The Hamiltonian for this control problem is
\begin{align}\label{eq:Unconstrainedham}
\begin{split}
    H=\;&L(x_\mathrm{1}(t),x_\mathrm{2}(t),u(t),t)\\&+\lambda_\mathrm{1}(t)\cdot \dot{x}_\mathrm{1}(x_\mathrm{1},u(t),t)\\&+\lambda_\mathrm{2}(t)\cdot \dot{x}_\mathrm{2}(x_\mathrm{2},u(t),P_\mathrm{em}(t)),  
\end{split}
\end{align}
with 
\begin{align}
\begin{split}
 L(x_\mathrm{1}(t),x_\mathrm{2}(t),u(t),t)=\dot{J}_\mathrm{TCO}=&\\ \mathcal{J}_{\mathrm{b,hp}}\cdot\dot{\delta}_\mathrm{deg,hp}+\mathcal{J}_{\mathrm{b,he}}\cdot\dot{\delta}_\mathrm{deg,he}&+\mathcal{J}_{\mathrm{Q}}\cdot (P_{\mathrm{ec,hp}}+P_{\mathrm{ec,he}}),  \label{eq:TCOinstant}
\end{split}
\end{align}
or in case of minimal energy consumption
\begin{align}
L(x_\mathrm{1}(t),x_\mathrm{2}(t),u(t),t)=\dot{J}_\mathrm{E}=&P_{\mathrm{ec,hp}}+P_{\mathrm{ec,he}}. \label{eq:econsinstant}
\end{align}
We do not set constraints on the terminal energy of either battery.
% The Hamiltonian \labelcref{eq:Unconstrainedham} does not have a penalty function for the constraints. The constraints are governed by the feasible domain $\mathcal{U}$ of the control variable $u(t)$.
% The feasible domain is determined by \cref{eq:PDCconstraints}. 
% $L(x_\mathrm{1}(t),x_\mathrm{2}(t),u(t),t)$ is the gradient of the optimization objective, which is the time derivative of \cref{eq:TCO,eq:Econs}. The gradient of both \cref{eq:TCO,eq:Econs} is presented in \cref{eq:TCOinstant,eq:econsinstant}.
The co-states, $\lambda_\mathrm{1}$ and $\lambda_\mathrm{2}$, describe the influence of the dynamics on the optimal solution. The co-state dynamics are obtained by taking the partial derivative of the Hamiltonian w.r.t.\ the states
\begin{align}\label{eq:costatedyn}
    \dot{\lambda}(t)=\left. -\pdv{ H}{ x}\right|_{u_\mathrm{t}}.
\end{align}
The equations that define the Hamiltonian -- the cost function and system dynamics -- are all dependent on the systems states $x(t)$ through the dependency of $V_\mathrm{oc}(SoC)$ and $R(SoC)$ on $SoC$. However, the influence of the $SoC$ on both $V_\mathrm{oc}(SoC)$ and $R(SoC)$ are small during the drive cycle\cite{Onori2016}. Therefore, we obtain
\begin{align}\label{eq:costatedynzero}
\renewcommand{\arraystretch}{1}
    \dot{\lambda}(t)\approx \begin{bmatrix}
0\\0        
             \end{bmatrix},
\end{align}
and the co-states are only dependent on the initial estimated costate $\lambda_0$. All system constraints are governed by the feasible domain $\mathcal{U}$ of the control parameter. Since there are no constraints on the states, we define both co-states: $\lambda_1=0$ and $\lambda_2=0$. Finally, we can solve the power-split by finding the optimal $P_\mathrm{dc}$ at every time step for which~\eqref{eq:TCOinstant} or~\eqref{eq:econsinstant} are minimized.
\par
\subsubsection{Optimal sizing problem}
The sizing problem concerns the scaling of both the high-power battery and the high-energy battery. 
We define $\gamma$ as the scaling parameter between the capacities of the batteries. The scaling parameter $\gamma\in [0,1]$ defines the energy content distribution through
\begin{subequations}\label{eq:gammadef}
\begin{align}
    \mathcal{E}_\mathrm{hp}&=\gamma\cdot \mathcal{E}_{\mathrm{tot}},\\
    \mathcal{E}_\mathrm{he}&=(1-\gamma)\cdot \mathcal{E}_{\mathrm{tot}},
\end{align}
\end{subequations}
where $\mathcal{E}_{\mathrm{tot}}=60$kWh is the designed energy capacity
We define the design problem with control parameter $\gamma\in [0,1]$ in \cref{prob:scaling} 
% \begin{table}[t]
% \caption{Constraint parameters}\label{tb:design param}
% \centering
% {\tabulinesep=0.5mm
% \begin{tabu}{l|l|l|l}
% \hline
% Symbol & variable& value& Unit\\ \hline
% $\mathcal{E}_\mathrm{design}$&design energy capacity& $60$& $\mathrm{{kWh}}$\\
% $V_\mathrm{design}$&design full voltage& $400$&$\mathrm{V}$\\
% $P_\mathrm{design}$&design max power& $239$&$\mathrm{kW}$\\
% $\Delta t_\mathrm{a,min}$&acceleration time requirement&$6.1$&$\mathrm{s}$
% \end{tabu}}
% \end{table}
\begin{prob}[Optimal sizing problem]\label{prob:scaling}
The relative sizing of the batteries is the solution of
\begin{subequations}
\begin{align}
&\!\min_\gamma & &J_\mathrm{TCO}(P_\mathrm{dc}^\ast,\gamma)\:\vee\: J_\mathrm{E}(P_\mathrm{dc}^\ast,\gamma)&\nonumber\\
& \textnormal{s.t. } &&P_\mathrm{em,max} \leq \eta_{\mathrm{dc}}\cdot P_{\mathrm{he,max}}+P_{\mathrm{hp,max}}&\\
&&&V_{\mathrm{design}}=V_\mathrm{\{he,hp\},{full}}&\\
&&& P_\mathrm{dc}^\ast \in\mathbf{PSC}(\gamma)\\
& & &\labelcref{eq:V_oc T}-\labelcref{eq:VDC},~\eqref{eq:gammadef}&\nonumber
\end{align}
\end{subequations}
\end{prob}
\noindent
where $P_\mathrm{em,max}$ is the maximum motor power and $V_\mathrm{design}$ is the designed maximum battery voltage. We design the total battery pack such that it is capable of delivering the maximum motor power.

The last constraint defines the number of battery cells in series $N_\mathrm{s}$ for both the batteries according to  
\begin{align}\label{eq:seriescalc}
    N_\mathrm{s,\{he,hp\}}&=\left\lceil \frac{V_{\mathrm{design}}}{V_\mathrm{c,\{he,hp\},{max}}}\right\rceil,
\end{align}
with the full capacity cell voltage, $V_\mathrm{c,{max}}$, extracted from datasheets. 
With the number of cells in series fixed by the constraint, the number of cells in parallel $N_\mathrm{p}$, is defined according to
\begin{subequations}\label{eq:parrallel}
    \begin{align}
    N_\mathrm{p,hp}&=\left\lceil \frac{\gamma\cdot\mathcal{E}_{\mathrm{tot}}}{N_\mathrm{s,hp}\cdot V_\mathrm{c,hp,{nom}}\cdot\mathcal{Q}_\mathrm{c,hp}}\right\rceil,\\[1ex]
    N_\mathrm{p,he}&=\left\lceil\frac{(1-\gamma)\cdot\mathcal{E}_{\mathrm{tot}}}{N_\mathrm{s,he}\cdot V_\mathrm{c,he,{nom}}\cdot \mathcal{Q}_\mathrm{c,he}}\right\rceil.
    \end{align}
\end{subequations}
We optimize the design parameter $\gamma$ by means of an exhaustive search for the feasible domain of $\gamma\in [0, 1]$.

\section{Discussion}\label{ch:discussion}
This section expands on the limitations and assumptions presented in this work. 
 First, we use a simplified battery model to be able to capture multiple existing batteries, solely based on data provided by manufacturers. We scale the battery characteristics linearly without considering losses between battery cell connections. The addition of a more complex battery model could be beneficial if a full design study would be considered. 
\par
Second, we disregarded temperature effects in the power-split control problem, since temperature control by varying the power-split is against the objective of the control problem. In future research, a controllable cooling system could be included, but this is considered to be beyond the scope for now.
\par
Third, we do not constrain the final battery $SoC$. This might lead to control strategies that are not sustainable for repeated operations but will not restrict the optimal solution. Adding a constraint for the final battery $SoC$ could limit the results as different battery sizing would require different constraints. For example, a charge-sustaining strategy would be appropriate when the high-power battery is small but would penalize a configuration where the battery size is more or less equal. We therefore encourage further research into appropriate control strategies. In our paper, we refrain from any specific strategy to avoid bias to a certain solution.  
\par 
% The results of the optimization are subject to the parameters presented in our research.
% The reality is that these parameters, especially the costs, are not static. Energy prices have a very complex combination of dependencies. It is influenced by various socio-economic factors and many other influences, too.  The prices of the battery cells are also highly dependent on demand and future developments. 
% The battery chemical and electrical parameters are also subject to future developments but are not as prone to vary as much as the prices.  Therefore, we consider the conclusions of the cost optimization only within the domain we have presented in our research.  
% \par 
Finally, the research we perform is conceptual and derived from validated simulation methods. 
A physical test setup for one or multiple solutions presented would be a logical continuation of our research. 
\section{Results}\label{ch:resulst}
In \cref{ch:methodology}, we introduced the two optimization goals for three different battery combinations. This section analyses the results for the minimum energy consumption scenario, leaving the TCO optimization to be included in a future extension.
We simulate all three combinations of chemistries, NCA-NMC, NCA-LFP and NCA-LTO, for different capacity distributions determined by $\gamma \in [0, 1] $. For the different chemistries, we use the cells listed in Table~\ref{tb:cell types} with their respective performance characteristics shown in Table~\ref{tb:cell performance}. We exclusively present the results within the feasible domain. The boundary line pictured in the results represents the minimum power constraint of the design problem.    

\begin{table}[t]
\caption{Cell chemistry information}\label{tb:cell types}
\centering
\resizebox{\columnwidth}{!}{\begin{tabular}{l|l|l}
% \hline
Cell Type                  & Chemistry & Manufacturer: Cellname      \\ \hline
High Energy                & NCA            & Molicel: INR-18650-M35A   \\ \hline
\multirow{3}{*}{High Power}& NMC            & Sony: US18650VTC4   \\ \cline{2-3} 
                            & LFP            & Lithium Werks: ANR26650M1B   \\ \cline{2-3} 
                            & LTO            & ELB/Yinglong: LTO32140   \\ 
\end{tabular}}
\end{table}

\begin{table}[t]
\caption{Cell performance summary}\label{tb:cell performance}
\centering
{\tabulinesep=0.3mm
\begin{tabu}{l|l|l|l|l|l}
% \hline
Cell chemistry&NCA& NMC&LFP&LTO&Unit\\ \hline
Energy density&260&173&109&77&$\mathrm{\frac{Wh}{kg}}$\\
Power density&521&2467&5211&1150&$\mathrm{\frac{W}{kg}}$\\
Cycle life&400 &500 &4000&20000&$\mathrm{cycles}$\\
\end{tabu}}
\end{table}

\begin{figure*}
\centering
\begin{subfigure}[b]{0.32\textwidth}
  \includegraphics[width=\linewidth]{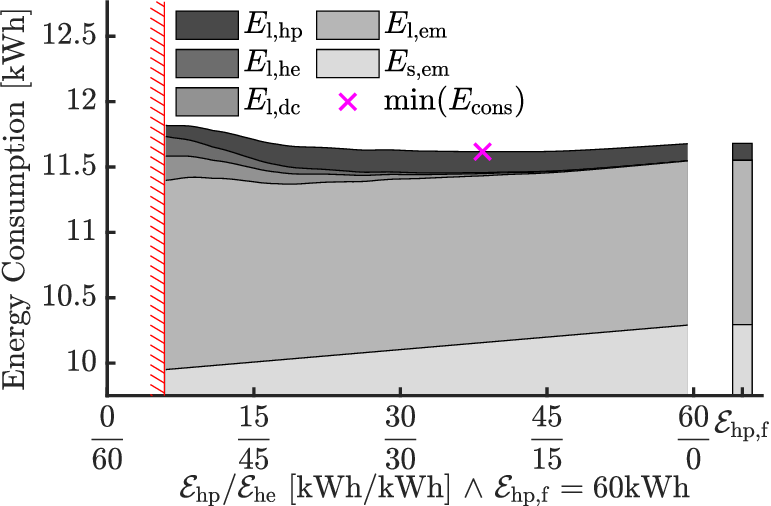}
  \caption{NCA-NMC energy breakdown}
  \label{fig:resultnmcE}
\end{subfigure}
\begin{subfigure}[b]{0.32\textwidth}
  \includegraphics[width=\linewidth]{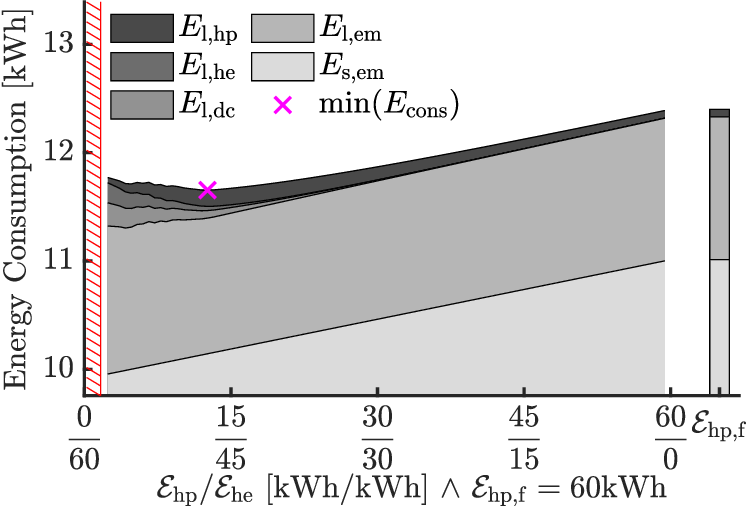}
  \caption{NCA-LFP energy breakdown}
  \label{fig:resultsLFPE}
\end{subfigure}
\begin{subfigure}[b]{0.32\textwidth}
  \includegraphics[width=\linewidth]{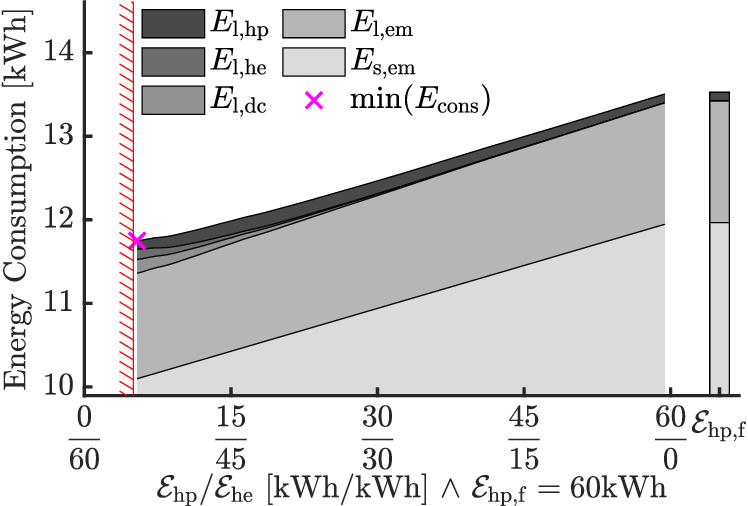}
  \caption{NCA-LTO energy breakdown}
  \label{fig:resultsLTOE}
\end{subfigure}
\begin{subfigure}[b]{0.32\textwidth}
  \includegraphics[width=\linewidth]{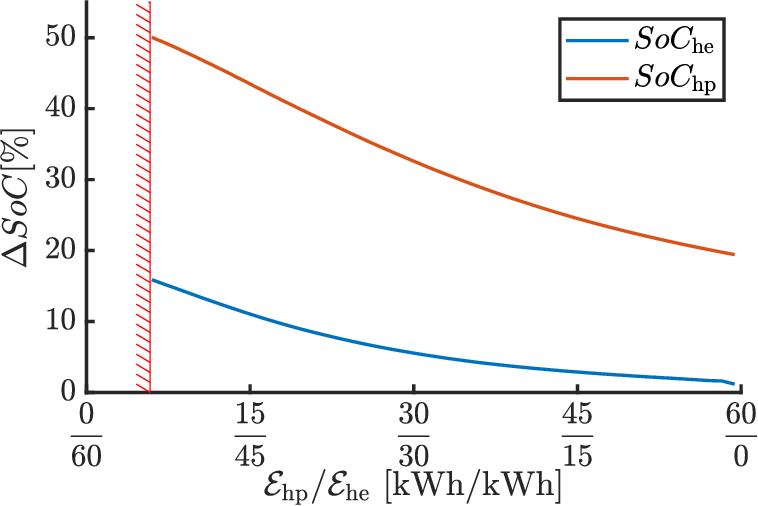}
  \caption{NCA-NMC $\Delta SoC$}
  \label{fig:SOCcostNMCE}
\end{subfigure}
\begin{subfigure}[b]{0.32\textwidth}
  \includegraphics[width=\linewidth]{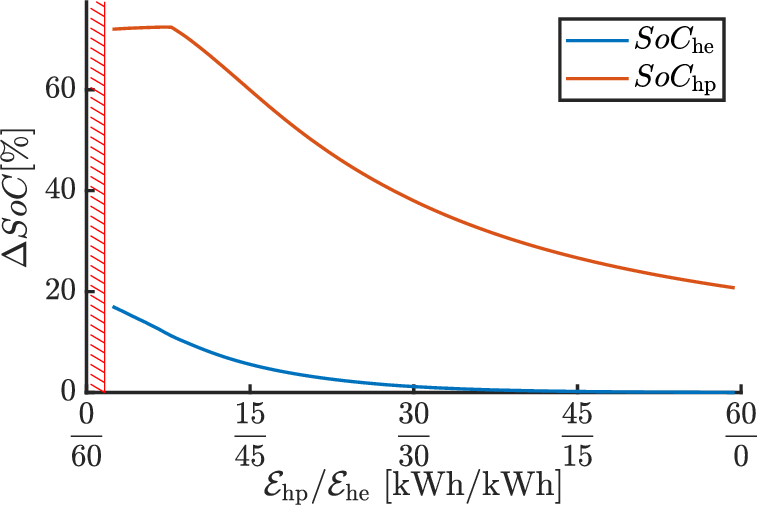}
  \caption{NCA-LFP $\Delta SoC$}
  \label{fig:SOCcostLFPE}
\end{subfigure}
\begin{subfigure}[b]{0.32\textwidth}
  \includegraphics[width=\linewidth]{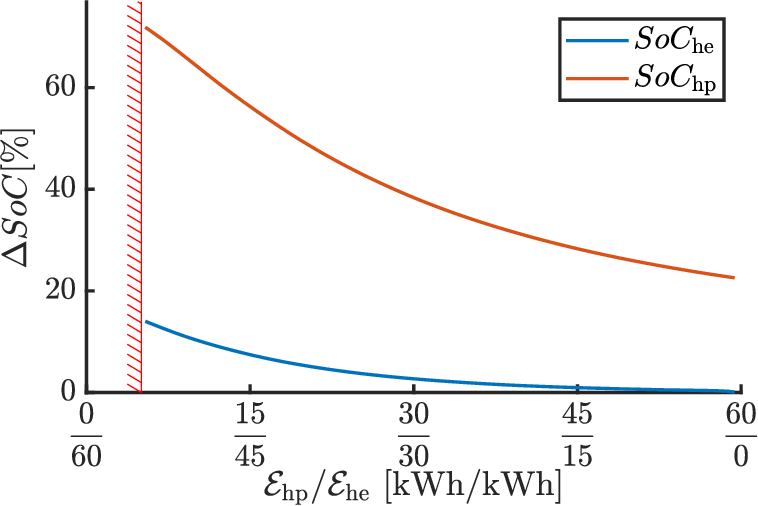}
  \caption{NCA-LTO $\Delta SoC$}
  \label{fig:SOCcostLTOE}
  \end{subfigure}
  \caption{Energy breakdown and $\Delta SoC$ for all feasible combinations for different capacity distributions between the high-power battery and the high-energy battery. The bar in Figures~\ref{fig:resultnmcE},~\ref{fig:resultsLFPE} and \ref{fig:resultsLTOE} represents the energy breakdown for a single cell battery constructed of only the power battery cell. The constraint represents the required maximum battery power of the total HESS.}
\label{fig:engresults}
\end{figure*}
%
% \subsection{Energy optimisation}
\par
To present the energy optimization results, we break the energy consumption down into the loss elements presented in \cref{tb:loss elements}.
\begin{table}[b]
\caption{Energy loss elements}\label{tb:loss elements}
\centering
{\tabulinesep=0.5mm
\begin{tabu}{l|l}
% \hline
Symbol & energy loss description\\ \hline
${J}_\mathrm{E}$& total energy consumption\\
$E_\mathrm{s,em}$&motor shaft energy consumption\\
$E_\mathrm{l,hp}$&energy loss in high-power battery\\
$E_\mathrm{l,he}$&energy loss in high-energy battery\\
$E_\mathrm{l,em}$&energy loss in motor\\
$E_\mathrm{l,dc}$&energy loss in DC-DC
\end{tabu}}
\end{table}
The results with $J_\mathrm{E}$ as optimization goal are presented in \cref{fig:engresults}. 
We calculate the energy breakdown for all battery combinations and the $SoC$ used for each battery during the simulation. The constraint indicates the minimum fraction of high-power batteries that are required to satisfy the maximum power requirement $P_\mathrm{em,max}$ of the complete HESS. Figures~\ref{fig:resultnmcE} to \ref{fig:resultsLTOE} present the capacity distribution at which the energy consumption is minimal. The value for the minimal energy consumption can be found in \cref{tb:optiengvalues}.
\begin{table}[b]
\caption{Energy optimisation values}\label{tb:optiengvalues}
\centering
{\tabulinesep=0.7mm
\begin{tabu}{l|l|l}
% \hline
Battery cell combination & $\min(J_\mathrm{E})$ & capacity  $\mathcal{E}_\mathrm{hp}/\mathcal{E}_\mathrm{he}$\\ \hline
NCA-NMC&$11.62\,\mathrm{kWh}$&${38.4\,\mathrm{kWh}}\;/\;{21.6\,\mathrm{kWh}}$ \\
NCA-LFP&$11.65\,\mathrm{kWh}$&${12.6\,\mathrm{kWh}}\;/\;{47.4\,\mathrm{kWh}}$\\
NCA-LTO&$11.75\,\mathrm{kWh}$&${5.4\,\mathrm{kWh}}\;/\;{54.6\,\mathrm{kWh}}$
\end{tabu}}
\end{table}
The results show that the optimal energy solution for all battery combinations consists of a hybrid solution. 
The energy breakdown in Figures~\ref{fig:resultnmcE} to \ref{fig:resultsLTOE} show that the energy consumption at the motor shaft $E_\mathrm{s,em}$ increases as the size of the high-power battery increases. The only parameter of the battery sizing problem that influences the energy consumption at the motor shaft $E_\mathrm{s,em}$ is the vehicle weight, influenced by the battery weight. Since the specific energy of the NCA battery cell is the highest of all simulated battery cells, the motor shaft energy consumption will be lower when the NCA high-energy battery has a larger share of the total capacity.   
In the $\Delta SoC$ curves, the controller uses the high-power battery relatively more. This controller bias can be explained by comparing the computed efficiencies of the different electrical components at the optimal capacity distribution, presented in \cref{tb:optiengs}. The combined efficiency of the DC-DC and the high-energy battery is lower for all cases than the high-power battery efficiency. 
\begin{table}[b]
\caption{Efficiencies at optimal point }\label{tb:optiengs}
\centering
{\tabulinesep=0.5mm
\begin{tabu}{l|l|l|l}
% \hline
Efficiency &NCA-NMC& NCA-LFP& NCA-LTO\\ \hline
$E_\mathrm{t,hp}/E_\mathrm{ec,hp}$&$0.987$&$0.982$&$0.977$\\
$E_\mathrm{t,he}/E_\mathrm{ec,he}$&$0.992$&$0.989$&$0.986$\\
$\eta_\mathrm{dc}$&$0.980$&$0.980$&$0.980$\\
$\int_{0}^{t_\mathrm{f}} P_\mathrm{em}\;\mathrm{d}t/E_\mathrm{s,em}$&$0.886$&$0.886$&$0.884$\\
\end{tabu}}
\end{table}
\par
The optimal energy consumption, shown in \cref{tb:optiengvalues}, is the lowest for the NCA-NMC combination during the drive cycle. However, we note that the difference between the optimal energy consumption of all combinations is small, i.e., about $1\%$. \par
\begin{figure}[t]
\centering
\includegraphics[width=0.95\columnwidth]{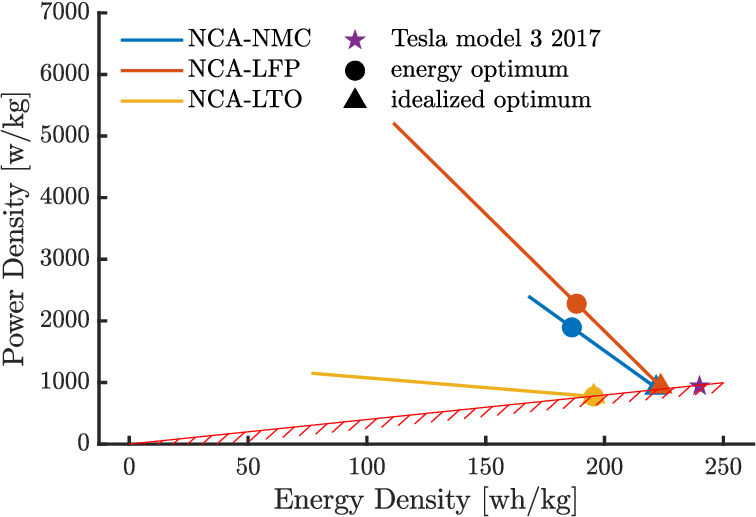}
\caption{Energy density vs power density of the investigated battery combinations for all feasible capacity distributions and the Tesla Model 3 2017 battery cell \cite{Batemotesla} with power constraint boundary line.}\label{fig:EvsPdens}
\end{figure}

In~\cref{fig:EvsPdens}, we present the simulated batteries in an energy density versus power density diagram to compare consumption results further.
The energy density increases with the size of the high-energy battery, and the power density increases with the capacity of the high-power battery. We indicate the power and energy density of the energy optimal results with a circle. The optimal results all have a similar energy density but a different power density. The similar energy density of the optimal solutions explains the small difference in energy consumption. For the NCA-LFP and NCA-NMC combinations, we observe that the optimal results do not maximize the energy density. We investigate this further by optimizing with an idealized lossless  DC-DC converter, indicated with a triangle in \cref{fig:EvsPdens}. In this case, the energy density is maximized, which demonstrates that the optimal solution capacity distribution forms a trade-off between reducing the DC-DC converter losses and increasing the energy density. This phenomenon indicates that increasing the DC-DC efficiency has a double advantage, since it reduces the losses and enables an optimal solution with a higher energy density. 
In summary, given the constraints, it is beneficial to construct a hybrid battery when optimizing for energy consumption. We can construct a battery with lower overall weight when introducing the high-energy battery while fulfilling the power constraint $P_\mathrm{em,max}$. 
% \par
% All solutions presented in \cref{fig:costresults,fig:engresults} have a larger $\Delta SoC$ of the high-power battery than the high-energy battery. This means that unless the batteries balance after driving to make the SoC equal for the next drive cycle, the control strategy is not a sustainable strategy for the complete. For this study, we do not consider a constraint for the SoC at the end of the cycle. We recommend further studies that explore what constraints on the SoC are beneficial for the optimal control of hybrid battery systems.  

\section{conclusion}\label{ch:conclusion}
In this paper, we studied the benefits of a hybrid energy storage system (HESS) in an electric vehicle when minimizing energy consumption. For the HESS, we considered a high-energy battery connected to the motor-inverter via a bidirectional DC-DC converter and a high-power directly connected to the motor-inverter, and varied the high-power battery chemistry between NMC, LFP and LTO. To decide the power-split between the two batteries, we designed an optimal control approach based on PMP, with the possibility of including additional constraints on the battery energy levels in future research.

% We introduce the optimization framework for two minimization goals: a cost of ownership minimization and an energy consumption minimization.
% \par
% The cost optimization shows that a single-cell LFP battery has the lowest cost.
% The optimal sizing and control are biased towards the battery type with the lowest initial production cost and degradation cost or a tradeoff between the two costs.
% We conclude that a HESS is not beneficial for reducing the cost of ownership in the research frame we established.
% \par
The results of the energy consumption optimization revealed a benefit for a dual battery solution. The NCA-NMC combination achieved the lowest energy consumption, providing the best trade-off between efficiency and weight. 
\par
In future work, we would like to investigate the influence of adding constraints on the terminal battery energy levels on the optimal sizing. Furthermore, we suggest further research into combined temperature and power control for hybrid batteries and the realization of a physical HESS for physical simulation purposes. 

\section*{Acknowledgment}
\noindent We thank Dr.~I.~New for proofreading this paper. This paper was partly supported by the NEON research project (project number 17628 of the Crossover program which is (partly) financed by the Dutch Research Council (NWO)).

\bibliographystyle{IEEEtran}
\bibliography{references.bib,MyCollection.bib}

\ifCLASSOPTIONcaptionsoff
  \newpage
\fi
% \usepackage[style=ieee]{biblatex}
% \addbibresource{MyCollection.bib}
% \addbibresource{references.bib}
\end{document}